# Toward a Science of Autonomy for Physical Systems: Defense

Ronald C. Arkin
arkin@cc.gatech.edu
Georgia Institute of Technology

Gaurav S. Sukhatme
gaurav@usc.edu
University of Southern California

## Once more unto the breach…

Militaries around the world have long been cognizant of the potential benefits associated with autonomous systems both in the conduct of warfare and in its prevention. This has lead to the declaration by some that this technology will lead to a fundamental change in the ways in which war is conducted, i.e., a revolution in military affairs (RMA) not unlike gunpowder, the long bow, the rifled bullet, the aircraft carrier, etc. Indeed the United States has created roadmaps for robotics with ever-increasing autonomous capability that span almost 40 years[2] These systems span air, sea, sea surface, littoral, ground and subterranean environments.

Why the interest? What advantages do autonomous systems afford the military? There are many, some of which include:

- Force multiplication where one warfighter may now be able to do the task of many, reducing the overall number of soldiers required for a military operation. This argues favorably both from an economic perspective as well as the ability to avoid conscription, a politically indelicate issue.
- Autonomous systems allow for an expansion of the battlespace, where operations involving greater persistence and longer endurance can be conducted over larger areas provide a strategic advantage.
- Extending the individual warfighter's reach allows the individual soldier to see further and strike further than would be otherwise available, increasing standoff distance from enemy threats.
- The net effect is a potential reduction in friendly casualties

---

[1] Contact: Ann Drobnis, Director, Computing Community Consortium (202-266-2936, adrobnis@cra.org).
For the most recent version of this essay, as well as related essays, please visit: cra.org/ccc/resources/ccc-led-white-papers
[2]*Unmanned Systems Integrated Roadmap: 2013-2028*, http://www.defense.gov/pubs/DOD-USRM-2013.pdf ; United States Air Force Unmanned Aircraft Systems Flight Plan 2009-2047, http://fas.org/irp/program/collect/uas_2009.pdf, accessed May 20, 2015.

There are serious societal and ethical concerns associated with the deployment of this technology that remain unaddressed. How can sufficient protection be afforded noncombatants? What about civilian blowback, where this technology may end up being used in policing operations against domestic groups? How can we protect the fundamental human rights of all involved? Considerable discussion is being conducted at an international level, including at the United Nations Convention on Certain Conventional Weapons (CCW) over the past two years, debating if and how such systems, particularly lethal platforms should be banned or regulated.

Part of the problem lies in the definition of autonomy – it is far from universally agreed upon. A high-level definition is a good starting point:

> *In its simplest form, autonomy is the ability of a machine to perform a task without human input. Thus, an "autonomous system" is a machine, whether hardware or software, that, once activated, performs some task or function on its own.[3]*

This will be our working definition for this position paper. We visit a broad range of DOD application areas that drive the use of this technology.

## DOD Application Area: A Sampler

Here, we present a small sampling of several representative areas of DoD relevant autonomous systems domains. There are many, many others – but this just serves to illustrate a small portion of ongoing research and associated needs.

Supply/resupply/logistics:

There are two primary areas of technology related to supply/resupply and logistics that are poised for significant impact: unmanned ground convoys and cargo drones. These robotic technologies are likely to make significant impact on large-scale resupply operations in both urban and non-urban settings. Large-scale aerial logistics[4] are needed in remote combat outposts with limited-to-no ground access. Such transportation can also effectively avoid ground-based threats. In both ground and aerial robotics, there is clear synergy with developments in the civilian sector in driverless cars and commercial drones. There are also challenges associated with airspace management in certain settings.

---

[3] *An Introduction to Autonomy in Weapon Systems*, P. Scharre and M. Horowitz, CNAS Working Paper, February 2015.
[4] DARPA's Aerial Reconfigurable Embedded System design (ARES) is an effort in this direction.

Reconnaissance

Over the past two decades, it has become clear that Army and Marine personnel actively seek and use small robots to perform reconnaissance and surveillance in battle (especially in urban environments). Such robots have been extensively used in Afghanistan and in Iraq. These robots have enjoyed considerable success in the field but they are yet to become "standard equipment" (accompanying doctrine in the Army and Marines on their use is not standardized). Since the cancellation of the Future Combat Systems program (2009) and the Small Unmanned Autonomous Ground Vehicle (SUGV) (2011) there is a gap in the future development and standardization of such robotic equipment for the ground forces. This is as much an acquisition challenge for the forces (since they primary use acquisition methods designed for tanks and planes) as it is a technology development challenge for the research community. The technology challenges in these vehicles are centered on autonomy in varied terrain, adapting to weather and seasonal changes, sliding-mode autonomy, multi-vehicle coordination, and the development of appropriate interfaces for operators.

There are currently 1000s of unmanned flying vehicles in use by the armed forces. Many are used for reconnaissance. Future challenges for these vehicles include the development of standards for their use in the forces. Technical challenges are centered in the areas of endurance/range, low-power sensing, disposability (fully bio-degradable vehicles?), and multi-robot coordination so that one warfighter may deploy a swarm of vehicles with little to no effort.

The development of robotic assets for underwater mapping missions, mine sweeps on the surface and reconnaissance missions in the air above water are all challenging areas that are of immediate interest to the Navy. The development of such assets is subject to the same challenges as aerial vehicles. Existing naval vessels also need significant retrofitting to accommodate robotics vehicles. It should be noted that the aquatic arena is extremely diverse encompassing deep ocean to littoral settings, harbors, and extending to lake and riverine settings.

Explosive Ordnance Disposal (EOD)

Robots have been used for EOD for several years. In this setting, they are primarily used for reconnaissance missions and delivering explosives for detonation. Both tasks are usually performed by an EOD technician teleoperating the robot. It is widely acknowledged that robots have saved many lives in this use case alone. There is tremendous scope in this area for robots with increased autonomy to dramatically reduce time on task – a key metric in this and many other military applications (including reconnaissance and scouting). There is also the potential for increased autonomy to allow a single technician to operate multiple vehicles (including mixed teams composed of aerial and ground vehicles).

Point man/Scout

A Point man or Scout robot is designed to keep the operator at a safe standoff distance while providing surveillance of urban structures, vehicles or other targets. Several such robots are commercially available and used by the Army. They have a significant amount in common with reconnaissance. Particular challenges with such robots arise in subterranean operations due to constraints on communication, lack of access to GPS, and the paucity of maneuvering room.

Prostheses/wounded warriors

Over the past two decades, there has been a revolution in prosthetic devices. Some of the most visible developments are in new materials for prostheses. These include carbon fiber, thermoplastic sockets and titanium. Other developments have led to the creation of robotic prostheses that (with embedded processors) are endowed with decision-making and control ability (e.g., to regulate joint resistance in the knee leading to stable walking). In addition, multiple prostheses (e.g., two legs) today exploit short-range communication (Bluetooth) to coordinate with each other. Modern prostheses are also capable of being controlled Myoelectrically (i.e., by electric signals from muscles) for greater precision. Finally, perhaps the most exciting area for the control of modern prostheses is directly by the amputee's mind.

## Concluding Remarks

The Department of Defense has accorded autonomy a high priority over the past several decades, not only focusing on development and deployment of these systems but also on the basic science of autonomy[5] underpinning this research. There is clear dual use for this military technology as has already evidenced from the outcomes of DARPA's Grand and Urban Challenges that have resulted in major advances for self-driving cars in the civilian sector. The same can be said for work on unmanned aerial vehicles (drones) now being considered for use in commercial package delivery systems for companies such as Amazon. As such, there are clear benefits from this research not only for national security but also in providing advances to benefit humanitarian, economic, and other sectors. DoD autonomous systems should continue to serve as a focal point for future research investment.

*For citation use*: Arkin R. & Sukhatme G. S. (2015). *Toward a Science of Autonomy for Physical Systems: Defense*: A white paper prepared for the Computing Community Consortium committee of the Computing Research Association. http://cra.org/ccc/resources/ccc-led-whitepapers/

*This material is based upon work supported by the National Science Foundation under Grant No. (1136993). Any opinions, findings, and conclusions or recommendations expressed in this material are those of the author(s) and do not necessarily reflect the views of the National Science Foundation.*

---

[5] http://www.onr.navy.mil/en/Media-Center/Fact-Sheets/Science-Autonomy.aspx